\documentclass[
twocolumn,
]{ceurart}

\usepackage{comment}
\usepackage{hyperref}
\begin{document}

\copyrightyear{2020}
\copyrightclause{Copyright for this paper by its authors.
  Use permitted under Creative Commons License Attribution 4.0
  International (CC BY 4.0).}

\conference{5th International GamiFIN Conference 2021 (GamiFIN 2021), April 7-10, 2021, Finland}

\title{Does gamification affect flow experience? A systematic literature review}


\author[1,3]{Wilk Oliveira}[%
orcid=0000-0003-3928-6520,
email=wilk.oliveira@usp.br,
]
\address[1]{Institute of Mathematics and Computer Science, University of São Paulo, São Carlos, São Paulo, Brazil}

\address[2]{Department of Information Systems, Brno University of Technology, Brno, Czech Republic}

\address[3]{Gamification Group, Faculty of Information Technology and Communication Sciences, Tampere University,
Tampere, Finland}

\author[2]{Olena Pastushenko}[%
orcid=0000-0002-0388-3115,
email=ipastushenko@fit.vut.cz,
]

\author[1]{Luiz Rodrigues}[%
orcid=0000-0003-0343-3701,
email=lalrodrigues@usp.br,
]

\author[1]{Armando M. Toda}[%
orcid=0000-0003-2681-8698,
email=lalrodrigues@usp.br,
]

\author[1]{Paula T. Palomino}[%
orcid=0000-0002-9730-2253,
email=paulatpalomino@usp.br,
]

\author[3]{Juho Hamari}[%
orcid=0000-0002-6573-588X,
email=juho.hamari@tuni.fi,
]

\author[1]{Seiji Isotani}[%
orcid=0000-0003-1574-0784,
email=sisotani@icmc.usp.br,
]
\begin{abstract}
  In recent years, studies in different areas have used gamification to improve users' flow experience. However, due to the high variety of the conducted studies and the lack of secondary studies (\textit{e.g.}, systematic literature reviews) in this field, it is difficult to get the state-of-the-art of this research domain. To address this problem, we conducted a systematic literature review to identify \textit{i)} which gamification design methods have been used in the studies about gamification and Flow Theory, \textit{ii)} which gamification elements have been used in these studies, \textit{iii)} which methods have been used to evaluate the users' flow experience in gamified settings, and \textit{iv)} how gamification affects users' flow experience. The main results show that there is growing interest to this field, as the number of publications is increasing. The most significant interest is in the area of gamification in education. However, there is no unanimity regarding the preferred method of the study or the effects of gamification on users' experience. Our results highlight the importance of conducting new experimental studies investigating how gamification affects the users' flow experience in different gamified settings, applications and domains.
\end{abstract}


\maketitle

\section{Introduction}
\label{sec:introduction}

\textit{Gamification} is ``the use of game elements (such as levels, points, badges, and others) in non-game contexts'' \cite{deterding2011game} and its techniques are used to enhance users experience in different areas, such as education \cite{dreimane2019gamification}, marketing \cite{noorbehbahani2019systematic}, healthcare \cite{muangsrinoon2019game}, and others. Different studies conducted over the past few years \cite{toda2017dark,koivisto2019rise,oliveira2020does} have shown that, if correctly applied, gamification can affect different types of positive psychological experiences (\textit{e.g.}, concentration, learning, and flow) in the users \cite{hamari2014does,koivisto2019rise,bai2020gamification}.

One of the main characteristics of the users experience that might be influenced by gamification is \textit{flow experience} \cite{oliveira2020does}, defined as a ``mental state in which a person performing some activity is fully immersed in a feeling of energized focus, full involvement, and enjoyment in the process of the activity'' \cite{csikszentmihalyi1975beyond,csikszentmihalyi2014validity,hong2019effect}. This is because one of the goals of gamification is to take users to a positive experience, some times, very related to flow \cite{deterding2011game}. Therefore, in recent years, different studies have proposed use various gamified tools to keep users in the state of flow during some activity, as well as, analyzing the effectiveness of different strategies to keep people in a flow experience \cite{huang2018create,zhou2019effect,kim2019impact}.

However, the results related to the effects of gamification on people's flow experience are still uncharted, and it is not possible to have an overview of the state-of-the-art. To face this challenge, in this study we conducted a systematic literature review to determine whether gamification techniques positively affect the users' flow state. To achieve this goal, we analyzed \textit{i)} what methods, tools and gamification elements have been mainly used in the published studies, \textit{ii)} what methods, techniques and tools have been used to analyze the flow experience in these studies, and \textit{iii)} what are the outcomes of the studies of flow experience in gamified settings.

The literature review results show that \textit{i)} there is an increasing interest to studying the effects of gamification on flow experience, \textit{ii)} education is the main domain addressed in the studies, \textit{iii)} there is no consensus on which methods and gamification elements to use to bring study participants to a flow experience, \textit{iv)} the questionnaire is the only method used to analyze the participant flow experience, and \textit{v)} the the effects of gamification on the flow experience are mixed (\textit{i.e.}, positive, negative, and neutral). Discussions section (Section \ref{sec:discussions}) reveals why more research in these areas is needed, based on the analysis of the existing studies.

Based on this premise, the main contribution of this paper is that it identified the most prevalent areas in the studies, which methods, techniques and game elements used in the studies, which methods are used to identify the users' flow experience and which are the possible effects of the gamification of the people's flow experience. Thus, as far as we know, this is the first study to provide an overview of the state-of-the-art in the field of gamification and Flow Theory (\textit{i.e.}, flow experience). The remaining of this paper is organized as follows. Section \ref{sec:method} presents the study method, Section \ref{sec:results} presents the results, discussion and limitations of the study, and Section \ref{sec:concluding-remarks} presents the study conclusion.

\section{Method}
\label{sec:method}

This systematic literature review aimed to identify the state-of-the-art of research on the effects of gamification on users' flow experience. To achieve this goal, we followed the well-known protocol proposed by Kitchenham \citep{kitchenham2004procedures}. It defines three general steps (planning, conducting, and documenting) to conduct secondary studies. Following this protocol, as the first step, we defined four research questions (RQ):

\begin{itemize}
    \item \textbf{RQ 1}: What gamification design methods have been used in the studies about gamification and Flow Theory?
    \item \textbf{RQ 2}: What gamification elements have been used in these studies?
    \item \textbf{RQ 3}: What methods have been used to evaluate the users' flow experience in gamified settings?
    \item \textbf{RQ 4}: How gamification affects users' flow experience?
\end{itemize}

Through the RQ 1 and RQ 2, it will be possible to have an overview of the main methods, tools, and gamification elements used in the studies that relate gamification and Flow Theory, allowing then to identify, for example, if there is a consensus concerning use a specific group of gamification elements, or if there is space for the use of new methods, tools or gamification elements in future studies in this field. Through RQ 3, it will be possible to identify how the participants' flow experience is analyzed in gamified settings, and consequently, as if there is a consensus about using a specific strategy to analyze the participants' flow experience. Finally, through RQ 4, it will be possible to get a general sense of the results of using gamification in the participants' flow experience. We believe that this set of questions will allow us to have a general notion of the state-of-the-art in studies that relate gamification and Flow Theory.

The next step of finding the primary studies, based on the selected protocol, is defining a ``search string''. Following different recent literature reviews \cite{koivisto2019rise,KLOCK2020102495}, we defined the search string based on the PICOC (\textit{Population, Intervention, Comparison, Outcomes, and Context}) method described by Kitchenham and Charters \cite{kitchenham2007guidelines}. Thus, the following PICOC was defined:

\begin{itemize}
    \item \textbf{Population}: studies that describe, apply or evaluate gamification to provide users' flow experience;
    \item \textbf{Intervention}: methods used to provide or evaluate users' flow experience in gamified systems;
    \item \textbf{Comparison}: not applicable, since the purpose of this study is to describe the state-of-the-art;
    \item \textbf{Outcomes}: most used methods to provide or evaluate users' flow experience in gamified systems;
    \item \textbf{Context}: studies in the field of gamification.
\end{itemize}

After applying the PICOC method, the generated search string was validated by a comparison with topics presented in recent studies in this field \cite{dos2018flow,koivisto2019rise,bai2020gamification,KLOCK2020102495}. We also validated our search string with two experts in the areas of gamification and Flow Theory. Thus, the final search string is: ``\textit{gamification AND (flow theory OR flow experience OR flow state)}''.

The next step was to define the sources. We conducted this step based on other recent literature reviews in the field of gamification and flow experience \cite{koivisto2019rise,bai2020gamification,KLOCK2020102495,dos2018flow}. Thus we defined four sources: ACM Digital Library\footnote{\url{https://dl.acm.org/}}; IEEE Xplorer\footnote{\url{https://ieeexplore.ieee.org/}}; Science Direct\footnote{\url{https://www.sciencedirect.com/}}; Springer Link\footnote{\url{http://link.springer.com/}}. In the next step, we defined inclusion (IC) and exclusion criteria (EC):

\begin{itemize}
    \item \textbf{IC1}: primary studies about gamification and flow theory; 
    \item \textbf{EC1}: secondary and tertiary studies; 
    \item \textbf{EC2}: redundant studies (written by the same authors and addressing the same topic\footnote{For these cases, we used the most recent studies})
    \item \textbf{EC3}: gray literature (non peer-reviewed studies).
\end{itemize}

Afterwards, we defined the data to be extracted from each selected study:

\begin{enumerate}
    \item \textbf{study  information} (reference, title, authors list, authors' country, authors' affiliations, source type (journal or conference), source, publishing year, and abstract (based on Oliveira \textit{et al}. \cite{dos2018flow})); \item \textbf{application domain};
    \item \textbf{method used to provide users' flow experience}; 
    \item used \textbf{gamification elements};
    \item \textbf{method used to identify users' flow experience};
    \item \textbf{Study results} (outcomes).
\end{enumerate}

The data collection process was conducted in October (2020) and was conducted by two experts in gamification and Flow Theory. They are also experts in conducting secondary studies. The experts studied the title, abstract and general metadata of all studies. Inspired by recent secondary studies \cite{KLOCK2020102495,dos2018flow}, doubts regarding the inclusion of studies (\textit{i.e.}, whether or not a study is relevant to our research questions) were discussed between the two researchers. A web system for the management of secondary studies (Parsif.al\footnote{\url{https://parsif.al/}}) was used to ease the process of the literature review (\textit{e.g.}, managing which studies were already analyzed, whether the studies were rejected, performing quality assessment and data extraction).

\section{Results}
\label{sec:results}

\begin{figure*}
\centering\includegraphics[width=0.8\linewidth]{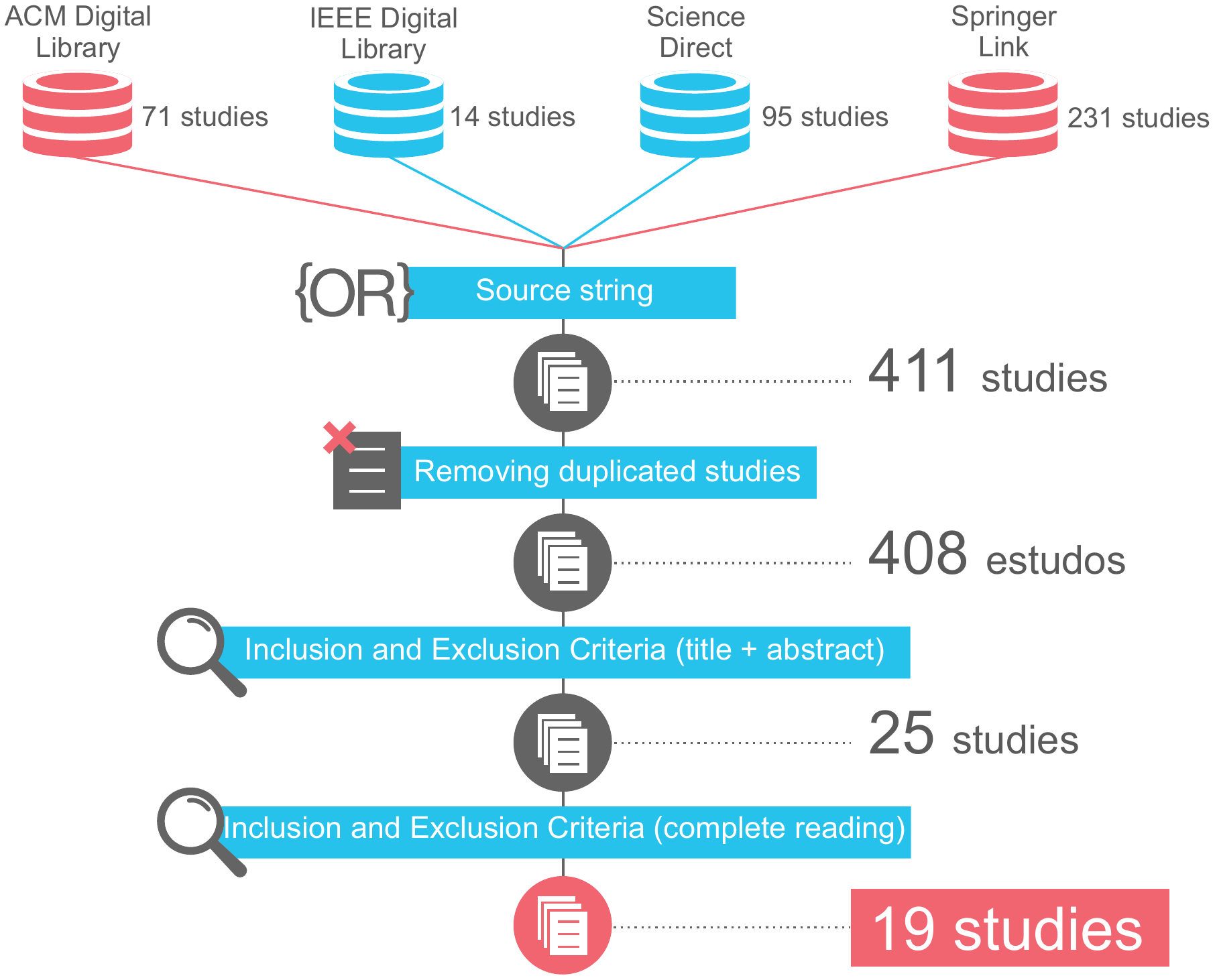}
\caption{Filtering}
\label{fig:filtering}
\end{figure*}

This section presents the literature review results, starting with the demographic information, and then answering the research questions afterward. In the first phase of the study, after executing the search string in the four research sources, 411 studies were found (ACM Digital Library = 71; IEEE Digital Library = 14; Science Direct = 95; and Springer Link = 231). Then, after removing the duplicate studies, 408 studies remained. After reading the studies' title and abstract, we found 25 studies, and finally, after the complete reading, we found 19 studies (ACM Digital Library = 3; IEEE Digital Library = 1; Science Direct = 4; and Springer Link = 11) that answered at least one of the selected research questions (and were corresponding to the inclusion criteria). \autoref{fig:filtering} presents an overview for the filtering process and \autoref{tab:list-of-selected-studies} presents the list of included studies. To facilitate the studies references throughout the text, we create `Ids' in the \autoref{tab:list-of-selected-studies}, which will be used whenever we are referring to a study.

\begin{table*}[]
\caption{List of selected studies}
\label{tab:list-of-selected-studies}
\begin{tabular}{lp{12cm}l}
\hline
\textbf{Id} & \textbf{Title}                                                                                                                           & \textbf{Ref.} \\ \hline
\hyperlink{cite.pastushenko2020methodology}{S01}         & A Methodology for Multimodal Learning Analytics and Flow Experience Identification within Gamified Assignments                                                  & \cite{pastushenko2020methodology}          \\ \hline
\hyperlink{cite.sehoon2020company}{S02}         & How a company’s gamification strategy influences corporate learning: A study based on gamified MSLP (Mobile social learning platform)    & \cite{sehoon2020company}          \\ \hline
\hyperlink{cite.palmas2020novel}{S03}         & A Novel Approach to Interactive Dialogue Generation Based on Natural Language Creation with Context-Free Grammars and Sentiment Analysis & \cite{palmas2020novel}          \\ \hline
\hyperlink{cite.quintas2020psychological}{S04}         & Psychological effects of gamified didactics with exergames in Physical Education at primary schools: Results from a natural experiment   & \cite{quintas2020psychological}          \\ \hline
\hyperlink{cite.shen2020motivates}{S05}         & What motivates visitors to participate in a gamified trip? A player typology using Q methodology                                         & \cite{shen2020motivates}          \\ \hline
\hyperlink{cite.chan2019understanding}{S06}         & Understanding the Effect of Gamification of Learning Using Flow Theory                                                                   & \cite{chan2019understanding}          \\ \hline
\hyperlink{cite.weng2019gamification}{S07}         & Gamification in Local Intangible Cultural Heritage Museums for Children: A Case Design                                                   & \cite{weng2019gamification}          \\ \hline
\hyperlink{cite.hogberg2019gameful}{S08}         & Gameful Experience Questionnaire (GAMEFULQUEST): an instrument for measuring the perceived gamefulness of system use                     & \cite{hogberg2019gameful}          \\ \hline
\hyperlink{cite.ilhan2019learning}{S09}         & Learning for a Healthier Lifestyle Through Gamification: A Case Study of Fitness Tracker Applications                                    & \cite{ilhan2019learning}          \\ \hline
\hyperlink{cite.antonaci2017gmoocs}{S10}         & gMOOCs – Flow and Persuasion to Gamify MOOCs                                                                                             & \cite{antonaci2017gmoocs}         \\ \hline
\hyperlink{cite.zhu2017improving}{S11}         & Improving Video Engagement by Gamification: A Proposed Design of MOOC Videos                                                             & \cite{zhu2017improving}         \\ \hline
\hyperlink{cite.korn2017strategies}{S12}         & Strategies for Playful Design When Gamifying Rehabilitation: A Study on User Experience                                                  & \cite{korn2017strategies}         \\ \hline
\hyperlink{cite.loos2017gamification}{S13}         & Gamification Methods in Higher Education                                                                                                 & \cite{loos2017gamification}         \\ \hline
\hyperlink{cite.roh2016goal}{S14}         & Goal-Based Manufacturing Gamification: Bolt Tightening Work Redesign in the Automotive Assembly Line                                     & \cite{roh2016goal}         \\ \hline
\hyperlink{cite.shi2014contextual}{S15}         & Contextual Gamification of Social Interaction – Towards Increasing Motivation in Social E-learning                                       & \cite{shi2014contextual}         \\ \hline
\hyperlink{cite.hamari2014measuring}{S16}         & Measuring flow in gamification: Dispositional Flow Scale-2                                                                               & \cite{hamari2014measuring}         \\ \hline
\hyperlink{cite.sepehr2013competition}{S17}         & Competition as an Element of Gamification for Learning: An Exploratory Longitudinal Investigation                                        & \cite{sepehr2013competition}         \\ \hline
\hyperlink{cite.ikehara2013combining}{S18}         & Combining Augmented Cognition and Gamification                                                                                           & \cite{ikehara2013combining}         \\ \hline
\hyperlink{cite.korn2012context}{S19}         & Context-Sensitive User-Centered Scalability: An Introduction Focusing on Exergames and Assistive Systems in Work Contexts                & \cite{korn2012context}         \\ \hline
\end{tabular}
\end{table*}

\begin{figure}
\centering\includegraphics[width=0.9\linewidth]{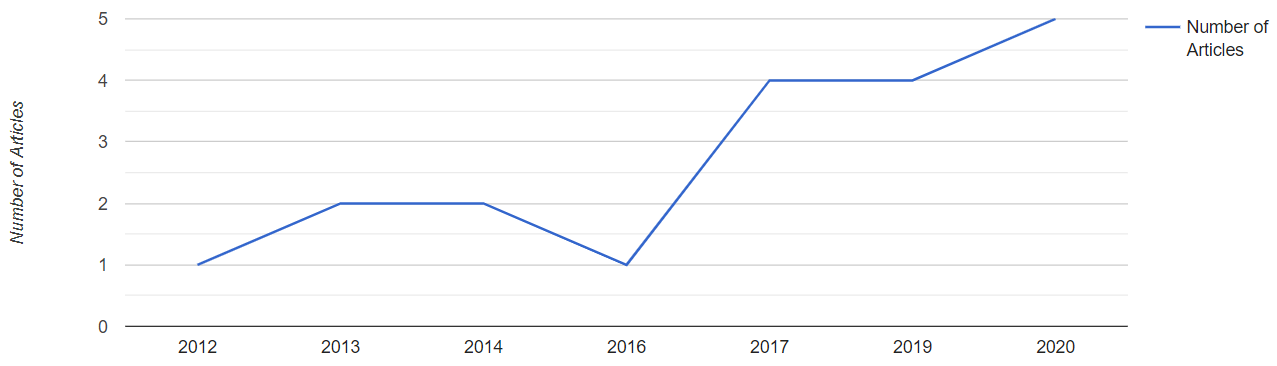}
\caption{Timeline of published studies}
\label{fig:timeline}
\end{figure}

\begin{figure*}
\centering\includegraphics[width=0.8\linewidth]{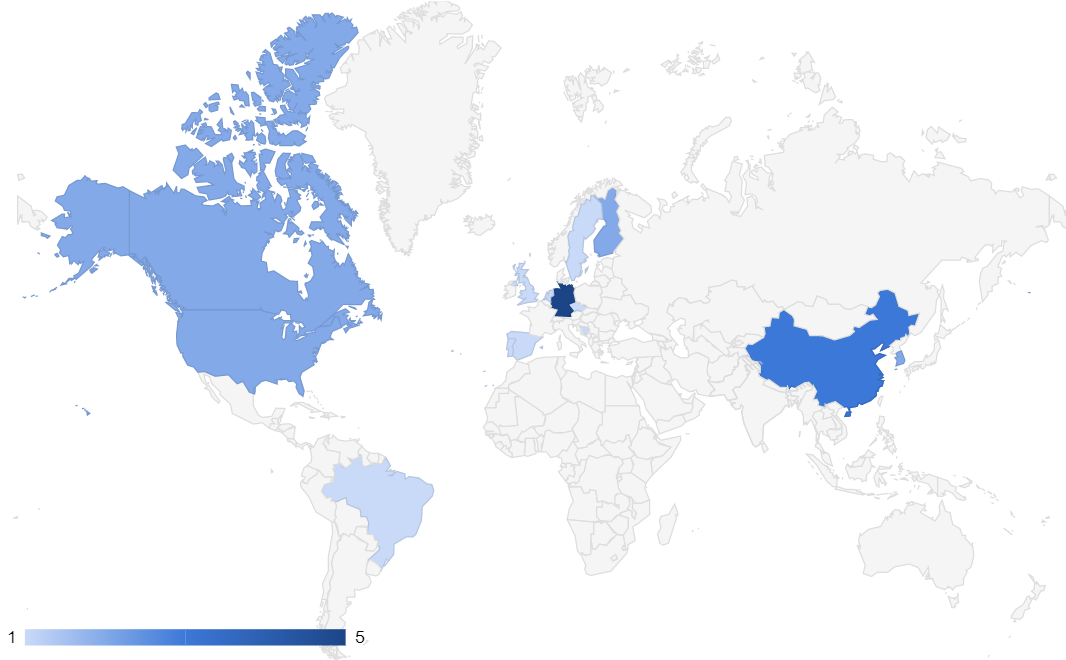}
\caption{Publications map}
\label{fig:publications-map}
\end{figure*}

The first study was published in 2012, and the majority of the studies were published in 2020. \autoref{fig:timeline} shows the number of publications over the years. Overall, the results indicate that there is a growing interest in researching the topic. In total, 64 different researchers, from 14 countries (Germany (five studies), China (three studies), Republic of Korea, Finland, USA, Canada (two studies each), Czech republic, Sweden, Spain, Netherlands, Brazil, United Kingdom, Bosnia and Herzegovina, Portugal (one study each)) contributed to 19 selected studies, and only one author participated in more than one study. \autoref{fig:publications-map} presents heat-map of publications according to the number of publications.

\begin{figure}
\centering\includegraphics[width=0.9\linewidth]{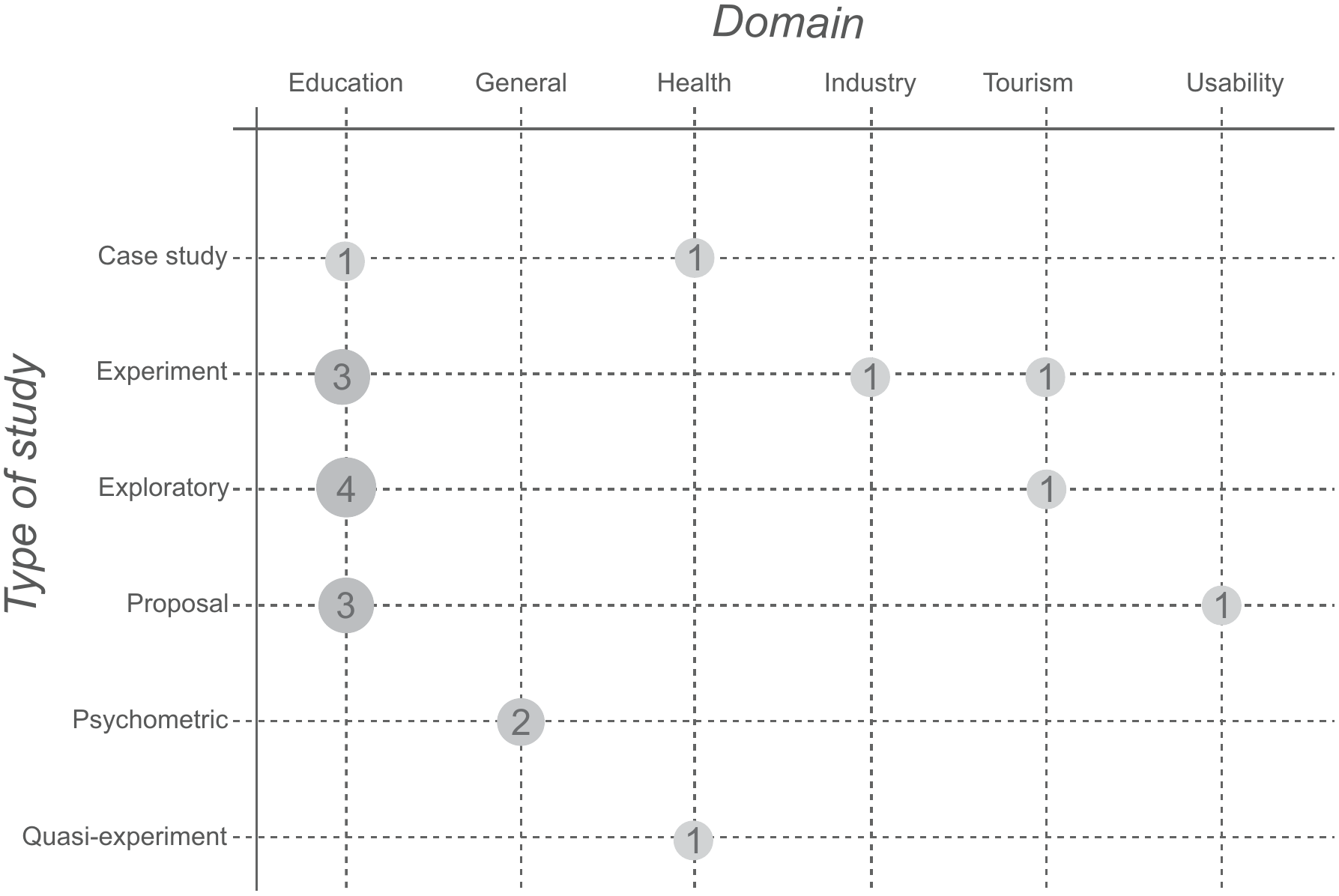}
\caption{Type of studies and domain}
\label{fig:type}
\end{figure}

From the 19 selected studies, five are exploratory studies (\hyperlink{cite.sehoon2020company}{S02}, \hyperlink{cite.palmas2020novel}{S03}, \hyperlink{cite.shen2020motivates}{S05}, \hyperlink{cite.loos2017gamification}{S13} and \hyperlink{cite.sepehr2013competition}{S17}), five are experimental studies (\hyperlink{cite.pastushenko2020methodology}{S01}, \hyperlink{cite.quintas2020psychological}{S04}, \hyperlink{cite.chan2019understanding}{S06}, \hyperlink{cite.weng2019gamification}{S07} and \hyperlink{cite.roh2016goal}{S14}), four are proposals (\hyperlink{cite.antonaci2017gmoocs}{S10}, \hyperlink{cite.zhu2017improving}{S11}, \hyperlink{cite.ikehara2013combining}{S18} and \hyperlink{cite.korn2012context}{S19}), two are case studies (\hyperlink{cite.ilhan2019learning}{S09} and \hyperlink{cite.shi2014contextual}{S15}), two are psychometric studies (\hyperlink{cite.hogberg2019gameful}{S08} and \hyperlink{cite.hamari2014measuring}{S16}), and one is a quasi-experimental study (\hyperlink{cite.korn2017strategies}{S12}). Eleven from the 19 selected studies were conducted in the field of education (\hyperlink{cite.pastushenko2020methodology}{S01}, \hyperlink{cite.sehoon2020company}{S02}, \hyperlink{cite.palmas2020novel}{S03}, \hyperlink{cite.quintas2020psychological}{S04}, \hyperlink{cite.chan2019understanding}{S06}, \hyperlink{cite.antonaci2017gmoocs}{S10}, \hyperlink{cite.zhu2017improving}{S11}, \hyperlink{cite.loos2017gamification}{S13}, \hyperlink{cite.shi2014contextual}{S15}, \hyperlink{cite.sepehr2013competition}{S17}, \hyperlink{cite.ikehara2013combining}{S18}), two in the field of health (\hyperlink{cite.ilhan2019learning}{S09}, \hyperlink{cite.korn2017strategies}{S12}), two in the field of tourism (\hyperlink{cite.shen2020motivates}{S05}, \hyperlink{cite.weng2019gamification}{S07}), two in the field of general gamification (\hyperlink{cite.hogberg2019gameful}{S08}, \hyperlink{cite.hamari2014measuring}{S16}), one the field of industry (\hyperlink{cite.roh2016goal}{S14}) and one in the field of usability (\hyperlink{cite.korn2012context}{S19}). \autoref{fig:type} presents a relation between the domains and type of studies.

\textbf{RQ 1: What gamification design methods have been used in the studies about gamification and Flow Theory?}. Only four from the 19 selected studies did not present the method used to provide the flow experience (\hyperlink{cite.sehoon2020company}{S02}, \hyperlink{cite.shen2020motivates}{S05}, \hyperlink{cite.hogberg2019gameful}{S08}, \hyperlink{cite.hamari2014measuring}{S16}). The other 15 studies presented the used method. However, only two (\hyperlink{cite.pastushenko2020methodology}{S01}, \hyperlink{cite.korn2017strategies}{S12}) also presented the used tools. \hyperlink{cite.pastushenko2020methodology}{S01} and \hyperlink{cite.ikehara2013combining}{S18} used \textit{gamified assignments} (the gamified assignments used by \hyperlink{cite.pastushenko2020methodology}{S01} is called bombsQuery), \hyperlink{cite.palmas2020novel}{S03} used a \textit{gamified dialogue scene}, \hyperlink{cite.quintas2020psychological}{S04} used \textit{gamified exergaming}, S06 used a \textit{gamified educational system}, \hyperlink{cite.weng2019gamification}{S07} used a \textit{gamified mock-up}, \hyperlink{cite.ilhan2019learning}{S09} used a \textit{gamified fitness tracking application}, \hyperlink{cite.antonaci2017gmoocs}{S10} and \hyperlink{cite.zhu2017improving}{S11} used \textit{gamified MOOCS}, \hyperlink{cite.korn2017strategies}{S12} used a \textit{gamified rehabilitation system} (called HUMAC NORM Testing), \hyperlink{cite.loos2017gamification}{S13} used a \textit{gamified classrooms} (sometimes using tailored gamification resources), \hyperlink{cite.roh2016goal}{S14} used a \textit{gamified bolt tightening tool}, \hyperlink{cite.shi2014contextual}{S15} used a \textit{gamified social e-learning environment}, \hyperlink{cite.sepehr2013competition}{S17} used a \textit{gamified simulation app}, and \hyperlink{cite.korn2012context}{S19} used a \textit{gamified assistive systems}. 

\textbf{RQ 2: What gamification elements have been used in these studies?}. Seven from 19 studies did not specify the gamification elements used in the study (\hyperlink{cite.pastushenko2020methodology}{S01}, \hyperlink{cite.sehoon2020company}{S02}, \hyperlink{cite.palmas2020novel}{S03}, \hyperlink{cite.shen2020motivates}{S05}, \hyperlink{cite.hogberg2019gameful}{S08}, \hyperlink{cite.korn2017strategies}{S12} ans \hyperlink{cite.hamari2014measuring}{S16}). Based on the studies that reported the gamification elements used, we identified that 31 different elements were used in the studies. The gamification elements \textit{``Level''} and \textit{``Points''} were the most used, being used in five different studies, followed by the gamification elements \textit{``Badges''} and \textit{``Leaderboards''}, used in four different studies. The gamification element \textit{``Goal''} was used in three different studies. \autoref{tab:used-gamification-elements} presents an overview on the used gamification elements throughout the selected studies, while \autoref{tab:methods-tools-and-elements} organizes the methods, tools, and gamification elements used in each study.

\begin{table}[ht]
\centering
\caption{Used gamification elements}
\label{tab:used-gamification-elements}
\begin{tabular}{p{5.5cm}l}
\hline
\textbf{Gamification   elements}                                                                                                                                                                                                                                                                                  & \textbf{Count} \\ \hline
Level, Points                                                                                                                                                                                                                                                                                                     & 5              \\ \hline
Badges, Leaderboards                                                                                                                                                                                                                                                                                              & 4              \\ \hline
Goals                                                                                                                                                                                                                                                                                                             & 3              \\ \hline
Countdown, Rewards, Feedback,   Immediate feedback, Progress bar, Storytelling                                                                                                                                                                                                                                    & 2              \\ \hline
Achievements, Audio-visual   feedback, Audio-visual feedback, Avatars, Challenges, Clues, Collaboration,   Communication Channels, Community Features, Competition, Customization,   Documentation, Empowerment, Flexibility, Guilds, Minigames, Narrative,   Quests, Skill trees, Smooth Learning Curves, Starts & 1              \\ \hline
\end{tabular}
\end{table}

\begin{table*}[]
\caption{Design methods, tools, and gamification elements}
\label{tab:methods-tools-and-elements}
\begin{tabular}{lllp{5.5cm}}
\hline
\textbf{Study} & \textbf{Design method}        & \textbf{Tools}     & \textbf{Gamification elements}                                                                                                                                  \\ \hline
\hyperlink{cite.pastushenko2020methodology}{S01} & Gamified assignment           & bombsQuery         & -                                                                                                                                                               \\ \hline
\hyperlink{cite.palmas2020novel}{S03} & Dialogue scene                & -                  & -                                                                                                                                                               \\ \hline
\hyperlink{cite.quintas2020psychological}{S04} & Gamified exergaming           & -                  & Points, Badges and Leaderboards                                                                                                                                 \\ \hline
\hyperlink{cite.chan2019understanding}{S06} & Gamified educational system   & -                  & Immediate Feedback, Rewards and Points                                                                                                                          \\ \hline
\hyperlink{cite.weng2019gamification}{S07} & Gamified mock-up              & -                  & Level, Starts and Feedback                                                                                                                                      \\ \hline
\hyperlink{cite.ilhan2019learning}{S09} & Fitness tracking application  & -                  & Goals, Points, Levels, Progress Bars, Feedback, Documentation, Badges, Leaderboards, Time, Quests, Avatars, Storytelling and Community Features                 \\ \hline
\hyperlink{cite.antonaci2017gmoocs}{S10} & MOOC                          & -                  & Empowerment, Smooth Learning Curves, Communication Channels, Levels, Clues, Goal Indicators, Skill trees, Guilds, Storytelling, Points, Badges and Leaderboards \\ \hline
\hyperlink{cite.zhu2017improving}{S11} & Videos MOOC                   & -                  & Narrative                                                                                                                                                       \\ \hline
\hyperlink{cite.korn2017strategies}{S12} & Rehabilitation System         & HUMAC NORM Testing & -                                                                                                                                                               \\ \hline
\hyperlink{cite.loos2017gamification}{S13} & Gamified classrooms           & -                  & Points, Challenges, Collaboration, Minigames, Achievements, Leaderboard,   Levels, Extra rewards, Flexibility and Countdown                                     \\ \hline
\hyperlink{cite.roh2016goal}{S14} & Bolt tightening tool          & -                  & Audio-visual feedback, Progress bar and badges                                                                                                                  \\ \hline
\hyperlink{cite.shi2014contextual}{S15} & Social e-learning environment & -                  & Goals, Immediate feedback and Customization                                                                                                                     \\ \hline
\hyperlink{cite.sepehr2013competition}{S17} & Simulation app                & -                  & Competition                                                                                                                                                     \\ \hline
\hyperlink{cite.ikehara2013combining}{S18} & Gamified Activity             & -                  & -                                                                                                                                                               \\ \hline
\hyperlink{cite.korn2012context}{S19} & Assistive systems             & -                  & Level                                                                                                                                                           \\ \hline
\end{tabular}
\end{table*}

\textbf{RQ 3: What methods have been used to evaluate the users' flow experience in gamified settings?}. Eleven from the 19 selected studies illustrated some method for analyzing the flow experience (the following studies did not specify a method used to analyze the flow experience: \hyperlink{cite.ilhan2019learning}{S09}, \hyperlink{cite.antonaci2017gmoocs}{S10}, \hyperlink{cite.zhu2017improving}{S11}, \hyperlink{cite.korn2017strategies}{S12}, \hyperlink{cite.loos2017gamification}{S13}, \hyperlink{cite.shi2014contextual}{S15}, \hyperlink{cite.ikehara2013combining}{S18}, \hyperlink{cite.korn2012context}{S19}). Coincidentally, all studies that reported the method used, used the same method to analyze the students' flow experience (\textit{i.e.}, questionnaire). However, different questionnaires were used (see \autoref{tab:flow-experience-evaluation-tools}). Additionally, \hyperlink{cite.hogberg2019gameful}{S08} conducted a Psychometric study to validate a questionnaire to identify the GAMEFUL experience \cite{landers2019defining} (GAMEFULQUEST), which includes the flow experience with one of its dimensions. \hyperlink{cite.hamari2014measuring}{S16} conducted a Psychometric study to validate the flow state scale proposed and validated by Jackson and Eklund \cite{jackson2002assessing} for the gamification domain.

\begin{table}[]
\centering
\caption{Flow experience evaluation tools}
\label{tab:flow-experience-evaluation-tools}
\begin{tabular}{lp{5cm}}
\hline
\textbf{Studies} & \textbf{Tools}                                           \\ \hline
\hyperlink{cite.pastushenko2020methodology}{S01}, \hyperlink{cite.weng2019gamification}{S07}         & Short Flow State Scale-2 \cite{jackson2002assessing}                                 \\ \hline
\hyperlink{cite.sehoon2020company}{S02}              & Four questions adjusted by Agarwal and Karahanna \cite{agarwal1997role}  \\ \hline
\hyperlink{cite.palmas2020novel}{S03}              & Game experience questionnaire (GEQ) \cite{ijsselsteijn2013game}                       \\ \hline
\hyperlink{cite.quintas2020psychological}{S04}              & Dispositional Flow State Scale-2 \cite{jackson2002assessing}      \\ \hline
\hyperlink{cite.shen2020motivates}{S05}              & Q sorts questionnaire adapted from Van Exel and De Graaf \cite{van2005q}                                    \\ \hline
\hyperlink{cite.chan2019understanding}{S06}              & Flow questionnaire Csikszentmihalyi and Larson \cite{csikszentmihalyi2014validity}               \\ \hline
\hyperlink{cite.roh2016goal}{S14}              & Experience sampling method (ESM) \cite{larson2014experience}                         \\ \hline
\hyperlink{cite.sepehr2013competition}{S17}              & Multidimensional measurement Mahfouz and Guo \cite{mahfouz2011overview} \\ \hline
\end{tabular}
\end{table}

\textbf{RQ 4: How gamification affects users' flow experience?}. Ten from the 19 selected studies present some outcome about the effects of gamification on the users' experience (in the following studies, authors did not report the results:
\hyperlink{cite.weng2019gamification}{S07}, \hyperlink{cite.ilhan2019learning}{S09}, \hyperlink{cite.antonaci2017gmoocs}{S10}, \hyperlink{cite.zhu2017improving}{S11}, \hyperlink{cite.korn2017strategies}{S12}, \hyperlink{cite.loos2017gamification}{S13}, \hyperlink{cite.shi2014contextual}{S15}, \hyperlink{cite.ikehara2013combining}{S18}, \hyperlink{cite.korn2012context}{S19}). Overall, the results are mixed. On the one hand, the results of \hyperlink{cite.palmas2020novel}{S03}, \hyperlink{cite.quintas2020psychological}{S04}, \hyperlink{cite.shen2020motivates}{S05}, \hyperlink{cite.chan2019understanding}{S06}, and \hyperlink{cite.sepehr2013competition}{S17} indicated that gamification did not significantly affect the users' flow experience. On the other hand, \hyperlink{cite.pastushenko2020methodology}{S01} identified that gamification positively affected the users' flow experience (in general, the users' flow experience in the gamified settings was high = 4.4). \hyperlink{cite.sehoon2020company}{S02} demonstrated that \textit{Challenge} (gamification design proposed to encourage challenge among participants), \textit{Relationship} (gamification design proposed to encourage relationship among participants), and \textit{Usability} positively affected flow and flow affected the intention of continuous use. \hyperlink{cite.roh2016goal}{S14} identified that a gamified interface improved worker's flow experience compared to other interfaces. Finally, \hyperlink{cite.hogberg2019gameful}{S08} and \hyperlink{cite.hamari2014measuring}{S16} obtained good model fits to validate its scales (\textit{i.e.}, GAMEFUL questionnaire \cite{hogberg2019gameful} and FSS2 \cite{hamari2014measuring}). \autoref{tab:outcomes-comparison} presents a comparison analyzing the types of outcomes in terms of domain, type of study and gamification elements.

\begin{table*}[]
\caption{Outcomes comparison}
\label{tab:outcomes-comparison}
\begin{tabular}{p{2cm}p{2cm}p{2cm}p{2cm}p{2cm}p{2cm}}
\hline
\multicolumn{3}{l}{\textbf{Weak effect}}                                                                                                                                                                                & \multicolumn{3}{l}{\textbf{Strong positive effect}}                                                                                              \\ \hline
\textbf{Domain}                                             & \textbf{Type of study}                                         & \textbf{Gamification elements}                                                                                             & \textbf{Domain}                                   & \textbf{Type of study}                              & \textbf{Gamification elements}                             \\
Education (\hyperlink{cite.palmas2020novel}{S03}; \hyperlink{cite.quintas2020psychological}{S04}; \hyperlink{cite.chan2019understanding}{S06}; \hyperlink{cite.sepehr2013competition}{S17}) Tourism (\hyperlink{cite.shen2020motivates}{S05}) & Exploratory (\hyperlink{cite.palmas2020novel}{S03}; \hyperlink{cite.shen2020motivates}{S05}; \hyperlink{cite.sepehr2013competition}{S17}) Experiment (\hyperlink{cite.quintas2020psychological}{S04}; \hyperlink{cite.chan2019understanding}{S06}) & Points (\hyperlink{cite.quintas2020psychological}{S04}, \hyperlink{cite.chan2019understanding}{S06}), Badges (\hyperlink{cite.quintas2020psychological}{S04}), Leaderboards (\hyperlink{cite.quintas2020psychological}{S04}), Immediate feedback (\hyperlink{cite.chan2019understanding}{S06}), Rewards (\hyperlink{cite.chan2019understanding}{S06}), Competition (\hyperlink{cite.sepehr2013competition}{S17}) & Education (\hyperlink{cite.pastushenko2020methodology}{S01}; \hyperlink{cite.sehoon2020company}{S02}), Industry (\hyperlink{cite.roh2016goal}{S14}) & Experiment (\hyperlink{cite.pastushenko2020methodology}{S01}; S14), Exploratory (\hyperlink{cite.sehoon2020company}{S02}) & Audio-visual feedback, Progress bar, Badges (\hyperlink{cite.roh2016goal}{S14}) \\ \hline
\end{tabular}
\end{table*}

In summary, the results indicate that \textit{i)} there is a growing interest of the community in analyzing the effects of gamification on the users' flow experience; \textit{ii)} the types of studies are varied, but most studies are exploratory and quantitative; \textit{iii)} there is a predominance of research in the field of education; \textit{iv)} many gamified methods have been used to provide the flow experience, but there is no unanimity in the type of method (although most can be generally characterized as gamified systems); \textit{v)} several gamification elements have been used, however, \textit{Level}, \textit{Points}, \textit{Badges}, and \textit{Leaderboards} are predominant; \textit{vi)} a single method has been used to analyze the users' flow experience (\textit{i.e.}, questionnaires), but there is no consensus concerning which questionnaire to use; \textit{vii)} there is no unanimity regarding gamification's effect on the users' flow experience, although promising results are observed.

\subsection{Discussions}
\label{sec:discussions}

In this paper, we presented the results of a secondary study on the field of gamification and Flow Theory. The overall results of our systematic literature review showed that most studies were conducted in the area of education. This result corresponds with the results observed in different secondary studies carried out in the large area of gamification \cite{hamari2014does,koivisto2019rise,bai2020gamification} who also identified that education is the area with the largest number of studies on gamification. This observation highlights the importance of conducting research analyzing how gamification affects the flow experience of users in specific contexts where there are few to no studies (\textit{e.g.}, marketing, tourism, health).

The studies on gamification started to become popular in the year 2012 \cite{deterding2011game,hamari2014does,hamari2014measuring}. It was precisely in that year that the first study on the use of gamification in the one related to Flow Theory appeared (\hyperlink{cite.korn2012context}{S19}). This discovery indicates that since the first studies on gamification, it was already thought of as a mechanism capable of influencing the people's flow experience. However, the number of studies in this research domain started to increase from 2016. This result shows that there is a growing interest of researchers to investigate the effects of gamification on people's flow experience, as well as that there is a growth perspective in the number of studies in this field (see \autoref{fig:timeline}).

At the same time, most of the published studies so far are exploratory studies (see \autoref{fig:type}), which may indicate that the area is still looking for maturity. This can also indicate that the area is growing, generating the need to conduct new studies, especially experimental studies that can individually assess the effects of different methods, tools, frameworks, and gamification elements on people's flow experience.

Still, on the growth of the area and the need for further studies, the results of this systematic review identified that researchers from 19 countries (see \autoref{sec:results}) have been involved in research so far. However, continents like Africa and Oceania (see \autoref{fig:publications-map}) have not yet started to produce research in this regard. The result also indicates the importance of different countries to carry out new studies or replicate studies already carried out in other countries to analyze the data from different perspectives.

Not all studies have defined the methods and tools with the ultimate goal of providing a flow experience to participants (\textit{i.e.}, in some studies, it is not explicit what the purpose of the method or tools and whether it was designed to lead participants to a flow experience). However, it is clear that different methods and tools were used in the studies and that there is no agreement regarding the use of any specific method or tool in the studies about gamification and Flow Theory. Specifically about tools, because only two studies make the tools used explicit, it is not possible to have insights about the use of specific tools (\textit{e.g.}, intelligent tutors or virtual reality glasses) in the studies on this domain.

A large variety of gamification methods have been used to provide users' flow experience in gamified settings. This indicates that, in general, research on this subject is still in its initial phase and tend to grow in the coming years, as there is still no consensus on which methods to use, and which are the best gamification elements. Thus, we believe that new studies can be conducted by evaluating and comparing different methods and elements of gamification.

Although it is possible to identify which game elements were most used in the selected studies, the studies do not describe whether each element was specifically thought based on some dimension of the Flow Theory. The most used game elements in the studies, generally correspond to the same game elements used in the large gamification area (\textit{i.e.}, points, badges, and leaderboards). However, some unusual game elements (\textit{e.g.}, narrative and curves) were also used.

All studies that reported the method used to analyze the users' flow experience, used a questionnaire. This is explained by the fact that this method is the simplest (\textit{e.g.}, methods for automatic identification based on data analysis) \cite{dos2018flow}, and other methods are more expensive or that cannot be used massively (\textit{e.g.}, electroencephalograms or eye trackers) \cite{oliveira2019towardsDC}. On the other hand, it is surprising that there is no consensus regarding the questionnaire used, and that many studies use questionnaires that have not been empirically validated for the gamification domain (\textit{e.g.}, the questionnaire validated by Hamari and Koivisto \cite{hamari2014measuring}).

Some studies identified that gamification positively affected the flow experience of users, others demonstrated that the effects were negative, and in other cases, no effects were perceived. This result is also similar to the results found in other recent secondary studies on gamification \cite{toda2017dark,jarnac2020systematic}. This may have occurred due to two main facts: \textit{i)} studies analyzing the effects of gamification on the experience of people flow are still few, and \textit{ii)} different methods have been evaluated using different analysis techniques. In both cases, the results highlight the importance of conducting further research, assessing the effects of gamification on the users' flow experience.

At the same time, it is difficult to analyze what affected the participants' flow experience (\textit{e.g.}, method, tool, or element of gamification itself). This difficulty is because, in general, the studies do not report, for example, how each gamification element was used individually, nor was any moderating intervention used to analyze these effects. Therefore, it is still difficult to identify whether gamification alone affects people's flow experience, and what the exact effects of gamification are. Through our results, future studies may conduct meta-analyzes to specifically analyze these effects. At the same time, this result also draws attention to the importance that new experimental studies, following strict criteria, should be conducted to analyze, for example, the effects of specific gamification elements, on the experience of the flow of participants in different types of scenarios.

The results of our secondary study allow us to begin to understand an overview of the area, as well as to make room for new studies to start being produced based on our results. Thus, based on our results, future studies may seek to carry out quality analysis in the selected studies, that is, individually analyze each of the 19 studies on different aspects and thereby have an overview of the technical and scientific quality of the studies.

At the same time, to identify the individual effects of each method, tool, or game element in the participants' flow experience, a meta-analysis can be conducted and consequently increase and deepen the results obtained in our study. Through a meta-analysis, it will also be possible to reflect on future actions that can be taken to deepen the area and even propose a research agenda as well as in other recent studies in the area of gamification.

Finally, our results contribute directly to the area of gamification, and indirectly to areas such as information systems, human-computer interaction, and user-centered design, showing a state-of-the-art in studies that relate gamification and Flow Theory, showing that the area has grown in recent years, however, studies are still initial and deserve to be further developed, as well as, that there is a great opening of space for the development of new studies to propose and evaluate new gamified experiences to positively affect people's flow experience.

\subsection{Limitations}

Some inherent limitations to this type of study are present. In general, we use mechanisms to mitigate all perceived limitations. Initially, the search string for the study may not match all topics in the study area. To mitigate this limitation, we use a specific method for defining search strings (\textit{i.e.}, PICOC method) and validate the string with domain experts. The process of selecting studies and extracting data may have failed to include some studies. To mitigate this limitation, in both phases, the process was conducted by two experts. At the same time, in the case of disagreement, the experts debated each specific case. The term gamification is often confused with games, which may include studies that focus on games and not on gamified systems. To mitigate possible limitations in this regard, we always follow what has been defined by each study's authors. In other words, if the authors said it was a gamified application, we trust and consider it as gamification.

\section{Concluding Remarks}
\label{sec:concluding-remarks}

Understanding how gamification affects people' flow experience is a current challenge in gamification studies. In this paper, we present the results of a systematic literature review conducted to identify the state-of-the-art in the use of gamification related to Flow Theory. Following a well-known process for conducting secondary studies (\textit{e.g.}, systematic reviews), we found 19 studies that were analyzed to answer our research questions.

The results of our study show that several different methods and gamification elements have been used to get people into a flow experience. However, there is no consensus as to which methods and gamification elements should be used. There only is a consensus on using questionnaires to analyze the people's flow experience in gamified settings, to the detriment of other methods, such as interviews and interaction data analysis. Still, there is no consensus on which questionnaire to use. Finally, we identified that the results related to the effects of gamification in the users' flow experience are still inconsistent.

The results directly contribute to understanding the effects of gamification on the people's flow experience and decision-making in conducting future studies. As future studies, we suggest extending this systematic review by answering other research questions (deeper analysis of the existing research, quotation analysis, etc), analyzing the quality of the selected studies, investigating potential research biases, and proposing a research agenda based on the results. We also suggest that future studies invest in conducting meta-analyses considering the studies included in this systematic review.

\begin{acknowledgments}
  The authors would like to thank the grant provided by São Paulo Research Foundation (FAPESP), Projects: 2018/07688-1, 2020/02801-4, 2016/02765-2, and 2018/15917-0; CAPES and CNPq.
\end{acknowledgments}



\end{document}